\documentclass[%
 reprint,
 amsmath,amssymb,
 aps,
]{revtex4-2}

\usepackage{graphicx}
\usepackage{dcolumn}
\usepackage{bm}

\usepackage{graphicx}
\usepackage{newtxtext}
\usepackage{newtxmath}
\usepackage{natbib}
\usepackage{hyperref}
\usepackage[dvipsnames]{xcolor}
\usepackage{array}
\usepackage{physics}
\usepackage{geometry}
\usepackage{subfigure}
\usepackage{booktabs}
\usepackage{ragged2e}

\usepackage{tikz}
\usetikzlibrary{positioning}
\usepackage{mwe}

\usepackage{comment}
\hypersetup{
    colorlinks = true,
    urlcolor   = blue,
    citecolor  = black,
}

\newcommand{\RomanNumeralCaps}[1]
\linenumbers

\begin{document}


\title[Electrohydrodynamic Stresses from Hydrogen-Bond Network Dynamics in Water]{Electrohydrodynamic Stresses from Hydrogen-Bond Network Dynamics in Water}
 
\author{Pramodt Srinivasula}
\altaffiliation[Previously at ]{Department of Mechanical Engineering, TU Darmstadt, where part of this research was conducted.}
\affiliation{Electrosoft labs LLP, Mumbai, India}


\begin{abstract}
The resistance of hydrogen-bond networks to ambient flow in water produces viscoelectric stresses and contributes to electrostrictive pressure. Within Onsager’s nonequilibrium thermodynamic framework, a lattice-gas description of aqueous electrolytes is combined with a coarse-grained hydrodynamic representation of hydrogen-bonded molecular networks, where viscous dissipation is modeled through energetically equivalent Brownian entities. This formulation connects molecular structural information from experiments and molecular dynamics to a unified dipolar Poisson–Nernst–Planck–Stokes (dPNP–S) continuum theory, quantitatively reproducing the measured viscoelectric coefficient of \citet{jin2022direct} and contributions to electrostrictive pressure. These results identify a microscopic mechanism by which hydrogen-bond dynamics influence electrohydrodynamic flow.
\end{abstract}

\maketitle

\textit{Motivation}-- Water, though seemingly simple at the macroscale, exhibits complex behavior that reflects its intricate molecular structure. The hydrogen-bonding and dipolar interactions among the water molecules play a pivotal role in governing both dielectric and hydrodynamic responses, particularly under an external electric field, manifesting in phenomena such as dielectric saturation \citep{alvarez2023debye}, viscoelectric effect (VE)  \citep{jin2022direct, baer2019water} and electrostrictive (ES) pressure\citep{drude1894elektrostriktion,landau2013electrodynamics}.

The dielectric saturation influences electric double layer (EDL) behaviour via reducing permittivity from 80 to as low as 2\citep{fumagalli2018anomalously} during its evolution \citep{colosi2013anomalous} upto several micrometer distances from the electrode surface through hydrogen bonded (HB) networks \citep{zheng2006surfaces}. 
VE and ES are interpreted using the decades-old empirical formulations\citep{lyklema1961interpretation,landau2013electrodynamics} 
beyond the scope of their original experimental settings. For example, the VE coefficient value $10^{-15}\,\mathrm{m^{2}V^{-2}}$ for water is widely used in modelling, where it significantly influences the nanoscale hydrodynamics across a different range of temperature and salt concentration\citep{hsu2016electrokinetics,saurabh2023mathematical,mehta2025arresting}. 

Modern MD studies indicate that hydrogen-bonded water clusters associated with viscous behaviour comprise roughly 25 molecules at low shear, which gives an effective water cluster structural radius ($R_{\text{wc}}$) of approximately $0.5\pm0.1\,\mathrm{nm}$
\citep{gao2024structural,zong2016viscosity, maheshwary2001structure}. Whereas HB correlation dynamics can extend further over a distance $l_d=1.75\pm0.25$\,nm in the network \citep{elton2017origin} and manifest as a slower Debye-like timescale \citep{popov2016mechanism}. 
Beyond viscosity changes, \citet{zong2016viscosity} observed an approximately linear increase in activation free energy, $\delta(\Delta G)  \!\sim\!  600$ J/mol at 298 K for strong electric field variations $\delta |\mathbf{E}| \!\!\sim \!\! 1$ V/nm, attributed to hydrogen-bond network reorganization in a background medium of relative permittivity $80$.

Similarly, light- and neutron-scattering measurements \citep{elamin2013long} further showed that although no persistent molecular structures exist beyond a few molecular diameters, water exhibits a slower Debye-like relaxation over nanometer length scales. Similar observations were made in other polar liquids as well \citep{wang2014observation}. These observations are consistent with the dielectric spectrometry results of \citet{jansson2010hidden}, which identified a 4-5 orders slower Debye-like relaxation in pure water due to HB networks (despite some initial concerns regarding electrode insulation effects masking the results).

\begin{figure}[!b]
\vspace{-10pt}
\centering
\begin{tikzpicture}
\sbox0{\includegraphics[width=0.2\textwidth]{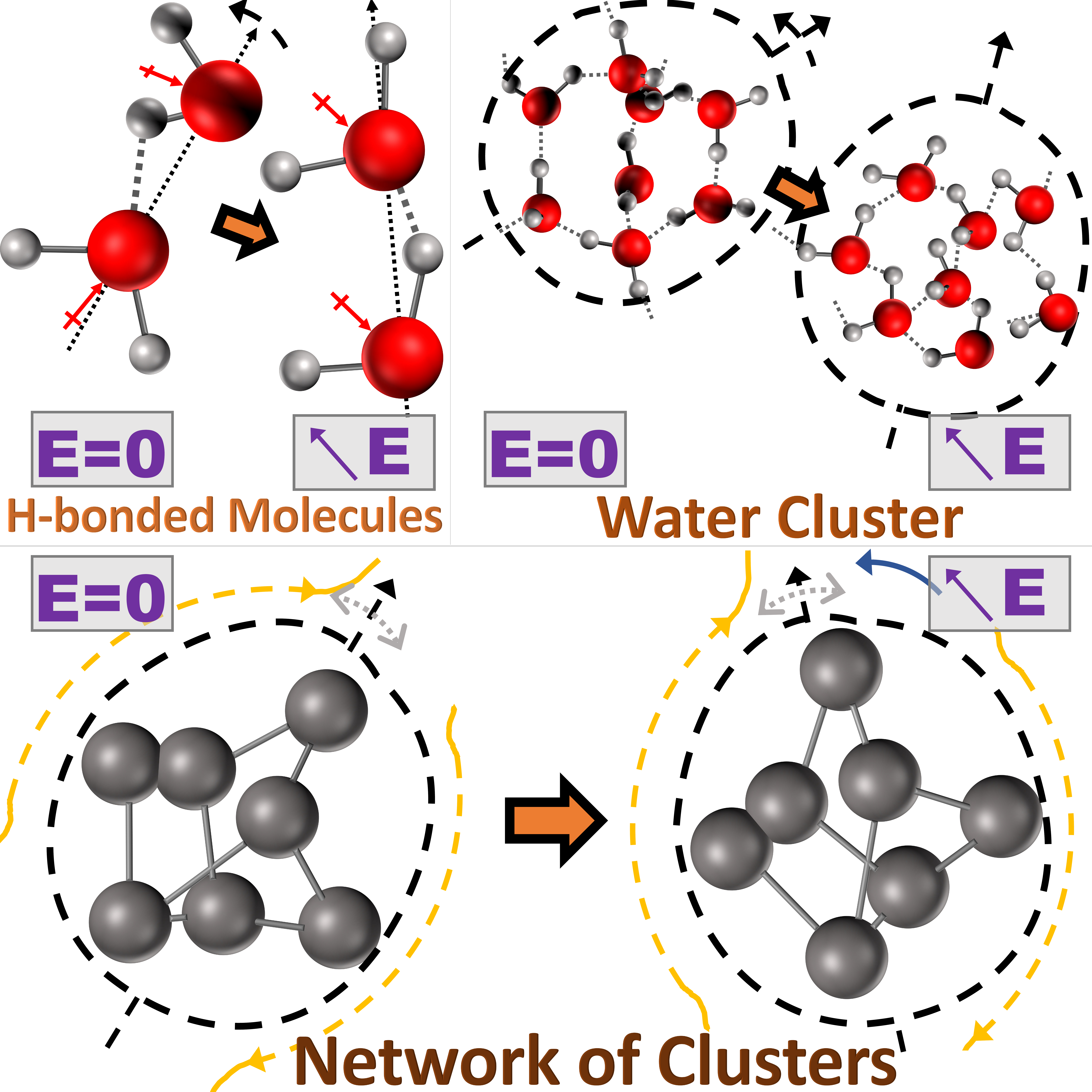} } 
  \node[above right,inner sep=0pt] at (0,0)  {\usebox{0}};
  \node[black] at (-0.05\wd0,1.0\ht0) {(a)};
\end{tikzpicture}
\begin{tikzpicture}
\sbox0{\includegraphics[width=0.2\textwidth]{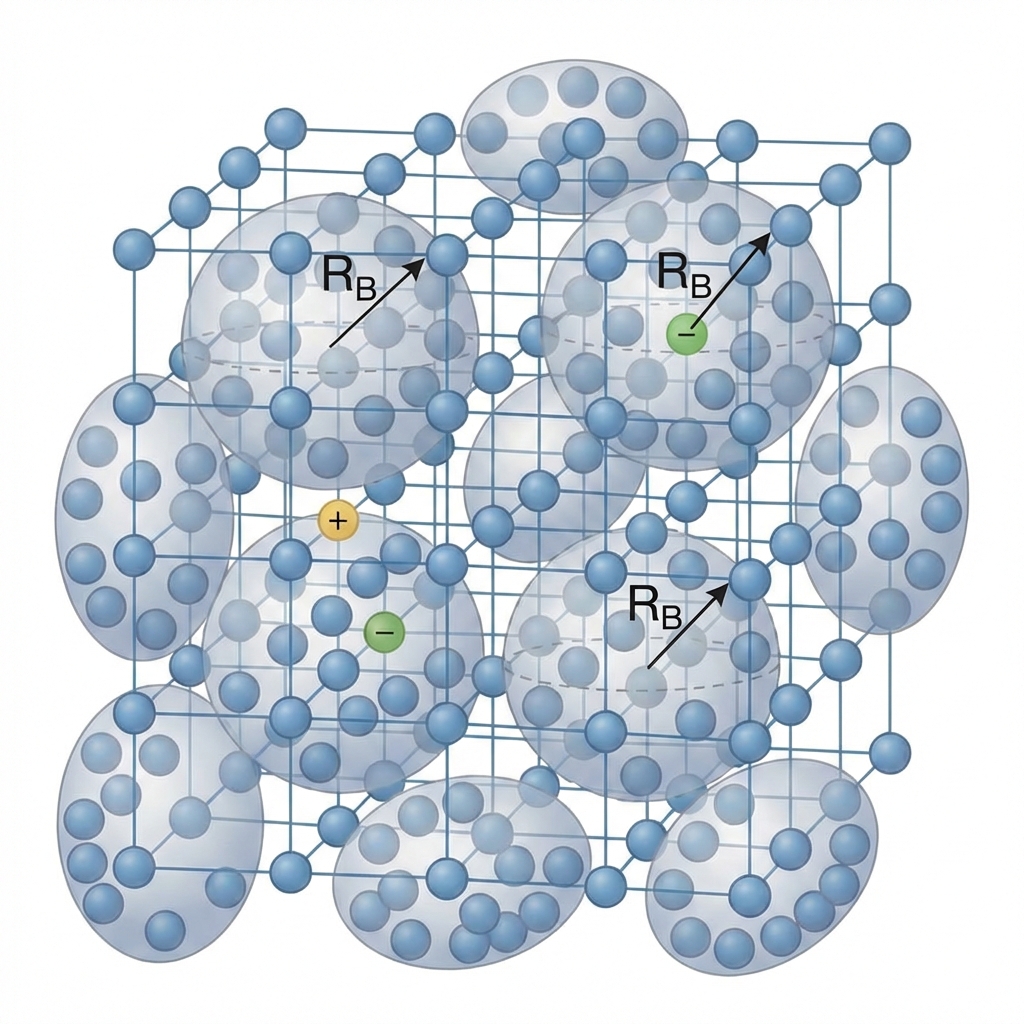} } 
  \node[above right,inner sep=0pt] at (0,0)  {\usebox{0}};
  \node[black] at (-0.05\wd0,1.0\ht0) {(b)};
\end{tikzpicture}
\caption{ Schematics: (a) Electric-field-induced hydrogen-bond restructuring across molecular, cluster, and network scales. Colored spheres denote oxygen (red), hydrogen (light gray), and water clusters (dark gray); solid and dashed bonds indicate covalent and HB correlations. Arrows indicate structural orientation (black) electric field \& induced rotation (blue), ambient flow (orange), and rotational diffusion (grey). (b) Coarse-grained Brownian-particle representation of the HB network embedded in a molecular lattice; sticks denote structural correlations between molecules (spheres). (Color online.)
} \label{fig:1_illustration}
\end{figure}

Despite these advances, a mechanistic continuum description of hydrogen-bond network dynamics in nanofluidic systems remains limited, particularly for realistic nanopores with lateral dimensions up to hundreds of nanometers involving transient electrokinetic processes \citep{he2011gate,emmerich2024nanofluidic}. 
This work introduces a coarse-grained hydrodynamic theory and captures the additional hydrodynamic stress contributions in the Stokes equation arising from the collective dynamics of the hydrogen-bond network.
\textit{Theory}-- Guided by molecular dynamics simulations and experiments, the slow collective dynamics of hydrogen-bond network segments are modeled as orientable Brownian entities embedded within the molecular lattice-gas description of the electrolyte. These entities represent coarse-grained segments of the hydrogen-bond network whose orientational dynamics generate additional dielectric and hydrodynamic stresses during electrical double-layer evolution in nanochannels of width $10\!-\!100$ nm.
Viscous dissipation associated with hydrogen-bond network dynamics is represented by identical Brownian particles with an equivalent coarse-graining radius $R_{B} \!= \!R_{\mathrm{wc}}\! +\! l_d\! =\! 2.25\,\mathrm{nm}$, uniformly filling the space with a concentration $c_B=R_B^{-3}$ m$^{-3}$. Hence, for water with molecular density $c_{s0}=55000 N_A$ m$^{-3}$ at 298 K (for density $990$ kg/m$^3$, Avogadro number $N_A$), each HB network segment analogous to a Brownian particle consists of $n_B\!\!\approx \!3020$ molecules.
The reorganization of water clusters, due to electric-field-induced realignment of dipolar molecules and the associated restructuring of the HB network, results in an effective reorientation of the virtual Brownian particles, as illustrated in Fig. \ref{fig:1_illustration}(a).

Consider an electric field $\mathbf{E}\!\!=\!\!|\mathbf{E}|\hat{\mathbf{E}}\!\!=\!\!-\nabla\psi$ imposed on the aqueous electrolyte. Based on the above data from\citet{zong2016viscosity}, reorientation of the representative Brownian particles along a strong electric field is interpreted as a linear field–energy coupling, $\delta (\Delta G) (n_B/N_A) \!\!\approx\!\!p_{0B}\delta(\mathbf{E})$. This defines a virtual, energetically equivalent Brownian particles dipole of moment $\mathbf{p}_B\!=\!p_{0B}\hat{\mathbf{p}}_B$ of constant magnitude $p_{0B}\simeq 633 $ D, characterized by a spatiotemporal mean-field orientation probability $f_{B}(p_{0B},\mathbf{E},\boldsymbol{\kappa},\mathbf{x},t)$. This decouples hydrogen-bond network energetics from the intrinsic molecular dipole contribution $\mathbf{\overline{p}}$, arising from individual dipoles and their interactions (per molecule), which determines the effective dipole moment per molecule $\mathbf{p}_{e}$ consistent with the zero-field relative permittivity $\varepsilon_{r0}$. Hence, $\mathbf{\overline{p}}\!=\!\mathbf{p}_{e}-\!\mathbf{p}_{B}/n_B$. Numerically, $p_{0B}$ is about one order of magnitude smaller than the summed molecular dipole moment of water within the corresponding hydrogen-bond network segment, with $|\mathbf{\overline{p}}|\sim 1.85-3$ D\citep{zong2016viscosity}) within the corresponding HB network segment. Thus a relevant parameter is identified as $\alpha_1\!=\!p_{0B}/(n_B|\mathbf{\overline{p}}|)\!\sim \!0.1$. 


The electric-field–induced reorientation of hydrogen-bond network segments is additionally modulated by thermal fluctuations, represented as orientational diffusion of the equivalent Brownian particles, and by hydrodynamic deformation through the local velocity-gradient tensor $\boldsymbol{\kappa}=\nabla\mathbf{v}$ and fluid viscosity $\eta$. Viscosity at zero electric field is $\eta_0 \approxeq  8.9\times 10^{-4}\,\mathrm{Pa\,s}$ at 298 K \citep{kaatze1989complex}. Consider a Jeffery's prolate spheroidal Brownian particle with aspect ratio (of major to minor axes lengths) $\Theta_B$. For weak geometric anisotropy, $\delta_B=(\Phi_B -1)\ll 1$, the difference in axial and equatorial rotational resistance coefficients scales linearly as $\sim 0.3\, \delta_B$ \citep{satoh2003introduction}. This correction is therefore subleading, and the rotational friction may be taken isotropic to leading order as,  $\zeta_B^r = 8\pi \eta_0 R_{B}^{\,3}\,F_{\mathrm{eq}}/\Theta_B^{\,2}$. 
{\setlength{\belowdisplayskip}{6pt}
\setlength{\abovedisplayskip}{6pt}
\begin{align}\label{eq:Perrin's factor}
    F_{\mathrm{eq}} \!&= \!\frac{4}{6}\frac{(1/\Theta_B)^2 - \Theta_B^2}{1\!-\!\left[\tanh^{-1}\!\left(\!\sqrt{1-\frac{1}{\Theta_B^2}}\!\right)\!\middle/\!\sqrt{1\!-\!\frac{1}{\Theta_B^2}}\!\right]\!(2-\frac{1}{\Theta_B^2}\!)}
\end{align}
}
is Perrin’s rotational friction factor \citep{koenig1975brownian}. Approximating its rotational diffusivity $D_{r,B}$ from the Stokes–Einstein relation $\beta D_{r,B}\zeta_B^r=1$, the first-order rotational relaxation time is $\tau_{r,B} = \zeta_{r,B }\beta/2$. Here $\beta$ is the inverse temperature parameter. Matching this to the experimentally observed Debye-like timescale of $29.5$\,ns for pure water at 298 K \citep{jansson2010hidden}, we obtain a small anisotropy $\delta_B= 0.0273$.

The coarse-grained representation is combined with the standard uniform molecular lattice gas model (Fig. ~\ref{fig:1_illustration}(b)), where they capture different degrees of freedom, at the network and molecular levels, respectively. Water molecules are modeled as spherical dipoles of constant magnitude moments $\mathbf{\overline{p}}=\overline{p}_0\hat{\mathbf{p}}$ with a mean-field orientational distribution $f(\overline{p}_0,\mathbf{E},\mathbf{x},t)$. The mean-field free-energy functional $\mathcal{F}$ for a lattice representing a dilute to moderately concentrated aqueous electrolyte (inter-ion, ion-water cluster interactions ignored) can be expressed in terms of spatio-temporal ion concentrations $c_\pm(\mathbf{x},t)$ with valencies $z_\pm$, solvent concentration $c_s(\mathbf{x},t)$, weighted angular averages $<(\cdot)>=(\int d\hat{\mathbf{p}}\, f\, (\cdot))/\int d\hat{\mathbf{p}} $ over the orientational degrees of freedom of the dipoles, $<(\cdot)>_B=(\int d\hat{\mathbf{p}}_B\, f_B\, (\cdot))/\int d\hat{\mathbf{p}}_B $ for Brownian clusters:%
\vspace{-12pt}
\begin{widetext}
\vspace{-10pt}
\begin{eqnarray}
\mathcal{F}\! &=& \!
\!\int \!\!d\mathbf{x} \Big[ \left(
\psi \rho_{i}\!-\frac{\varepsilon_0}{2}|\nabla\psi|^2
\!\right)\! +\! \frac{ 1}{\beta}\!\left(\! c_+\!\ln\!\left(\!\frac{c_+}{c_0}\!\right)\!+ c_-\!\ln\!\left(\!\frac{c_-}{c_0}\!\right)
\!+\! c_s\!\ln\!\left(\!\frac{c_s}{c_{s0}}\!\right)
\!\right) +\lambda_1(c_+\! +\! c_-\! +\! c_s \!\!-\! c_L) \Big]-\! \!\int \!\!\!d\mathbf{x}\,c_s\!\Big[ g \overline{p}_0^2 \nonumber\\
&& 
+
\gamma_c\!
\left<\mathbf{\overline{p}}\!\cdot\!\nabla \psi \right>\!+\! 
\left( \!\frac{<\ln f\!>\!- \lambda_2\!\left(\!<\!f\!>\!-\!1\!\right)}{\beta}\! \right)\!\Big]
+\!\int \!\!\!d\!\mathbf{x}\, c_B\!\left[\!
\left<\mathbf{p}_{B}\!\cdot\!\nabla \psi \right>_B \!+\! \left( \!\frac{<\!\ln f_B\!>_B\!- \lambda_{2B}\!\left(<f_B>_B\!-1\right)}{\beta}\!
\right)\right].
\label{eq:totalFreeEnergy}
\end{eqnarray}%
\vspace{-10pt}
\end{widetext}
\vspace{-12pt}
Here, $\varepsilon_0$, $e_0$ denote the permittivity of free space and the elementary charge; $g$, $\gamma_c$ are the solvent dipolar reaction-field and cavity parameters; $c_0$, $c_{s0}$ are the bulk concentrations of ions and solvent; and $\rho_{i}=e_0(z_+c_++z_-c_-)$ is the ionic charge density. The first two integrals correspond to finite ion effects \citep{kilic2007bsteric}, and dipolar solvent effects \citep{iglivc2010excluded,gongadze2012decrease} from the molecular lattice model, respectively. These represent electrostatic field interaction with the ions and the field's self energy; translational entropy of molecular species (crowding effects) with lattice-occupation constraint of constant total lattice space concentration $c_L=c_{s0}+2c_0$ enforced by Lagrange multiplier $\lambda_1(\mathbf{x})$; solvent dipolar interactions energy with reaction field, dipole–field coupling with cavity effect; and orientational entropy with normalization via $\lambda_2(\mathbf{x})$, in the order. The last integral corresponds to the equivalent Brownian HB network clusters incorporating the field-induced energy and orientational entropy with normalization via $\lambda_{2B}(\mathbf{x})$.

The standard Langevin-Boltzmann distribution of molecular dipole orientations subjected to an external electric field of nominal magnitude $E_0$, $f=(\alpha_2 \tilde{E}/sinh(\alpha_2  \tilde{E})) e^{\alpha_2 \hat{\mathbf{p}} \cdot \tilde{\mathbf{E}}}$, is obtained from $\partial \mathcal{F}/\partial f=0$  \citep{iglivc2010excluded}. Here $\tilde{\mathbf{E}}=\tilde{E}\hat{\mathbf{E}}$ is a normalized electric field with magnitude $\tilde{E}=|\mathbf{E}|/E_0$ and $\alpha_2=\gamma_c \overline{p}_{0}E_0\beta$ is a dimensionless number indicating relative magnitude of electric field influence on the molecular dipolar orientation.  
The standard Onsager's non equilibrium thermodynamic principles for electrostatics $\delta \mathcal{F}/\delta \psi=0$, and species chemical potentials $\mu_{\pm,s}=M_{\pm,s}+\lambda_1=\delta\mathcal{F}/\delta c_{\pm,s}$ with ion fluxes $J_{\pm,s}=-\beta D_{\pm,s} c_{\pm,s}\nabla\mu_{\pm,s}$ \citep{kilic2007bsteric}, gives the Poisson and advective Nernst-Planck equations,
{\setlength{\belowdisplayskip}{6pt}
\setlength{\abovedisplayskip}{6pt}
\begin{align}
    \label{eq:Poisson}
\!\!    -\varepsilon_0 & \nabla^2\psi\!=\!\rho_i\!+\!\rho_{s}, \,\,\, \, \rho_{s}\!=\!-\!\nabla\!\cdot\! \left( \mathbf{P}\!\left(\!1\!+\!\frac{\alpha_1}{\gamma_c}\!\frac{<\!\hat{\mathbf{p}}_B\!>_B}{<\!\hat{\mathbf{p}}\!>}\right)\!\right)\\
\dot{ c}_\pm\! =& D_\pm \Big[ z_\pm e_0\beta \nabla\! \cdot \!\left(c_\pm \nabla \psi \right) +  \nabla^2 c_\pm+   \nabla\! \cdot\! \left( \frac{c_\pm \nabla c}{c_L-c} \right)\nonumber\\
    & \!  +\!\alpha_2  \!\nabla \!\cdot\! \left( c_\pm (\nabla\tilde{E}) \mathcal{L} \right) \!+\!\beta \nabla\!\cdot \!( c_\pm \nabla \mu_s)\!\Big]-   \nabla\! \cdot\! (c_\pm \mathbf{v}),
\label{eq:NP}
\end{align}
}
where overdot denote time derivative. The solvent molecular and HB network effects reflect as the additional dipolar charge $\rho_{s}$ and the dipolar solvent-induced ionic flux (second line of Eq. \eqref{eq:NP}). 
Here $\mathcal{L}\!=<\!\hat{\mathbf{p}}\!\cdot\!\hat{\mathbf{E}}\!>\! =\! coth(\alpha_2\tilde{E})\!-\!(\alpha_2 \tilde{E})^{-1}$ is the Langevin function and $\mathbf{P}\!=\!\gamma_c c_s\! <\!\mathbf{\overline{p}}\!>$ is the polarization density vector with $\left< \mathbf{\overline{p}} \right>\!=\!p_{0} \mathcal{L} \mathbf{\hat{E}}$, parallel to $\mathbf{E}$. 

The molecular  level degrees of freedom in the lattice model equilibrate into fixed material parameters, such as a contribution towards the solvent viscosity $\eta_f$ (excluding HB effects). However, the contributions arising from the slower, collective dynamics of HB-network segments remain to be evaluated. 
%
%
The Onsager’s variational functional (or Rayleighian) $\mathcal{R}$ incorporating both molecular transport and mesoscale hydrogen-bond network dynamics is constructed as the sum of a dissipation functional $\Phi$ and the rate of change of free energy $\dot{\mathcal{F}}$ \citep{doi2011onsager}.
{\setlength{\belowdisplayskip}{6pt}
\setlength{\abovedisplayskip}{6pt}
\begin{align}
  \mathcal{R} \!
    =& \Phi \!+ \!\dot{\mathcal{F}}\!
    -\! \iint\! d\mathbf{x}d\hat{\mathbf{p}}_B\, 
        \lambda_3(\dot{\hat{\mathbf{p}}}_B\cdot\hat{\mathbf{p}}_B)
   \! +\! \!\int\!\! d\mathbf{x} \lambda_4(\nabla\cdot\mathbf{v}) \label{eq:Rayleighian}\\
   \!\!  \Phi =& 
    \int \!\!d\mathbf{x} \Big[\frac{\eta_f}{4}(\boldsymbol{\kappa}+\boldsymbol{\kappa}^T)^2\nonumber\\
        &
     +\! \frac{c_{B}\zeta_B^r}{4} 
    \!\left\langle 
       2(\dot{\hat{\mathbf{p}}}_B\! -\! \dot{\hat{\mathbf{p}}}_{f,B})^2
      \!  + \!\overline{\chi}
     (\hat{\mathbf{p}}_B\! \cdot \!\boldsymbol{\kappa}\cdot\hat{\mathbf{p}}_B)^2
    \right\rangle_{B} \Big]\label{eq:Dissip_Phi}\\
        \dot{\mathcal{F}}\!
   \! =& \!\!\int\!\! d\mathbf{x}\left(
        \mu_+ \dot{c}_+
       \! + \!\mu_- \dot{c}_-
       \! + \!\mu_s \dot{c}_s
       \! + \!\frac{\delta\mathcal{F}}{\delta f}\dot{f} \!+ \!\frac{\delta\mathcal{F}}{\delta f_B}\dot{f_B}
    \right)\!
\end{align}
}
Lagrange multipliers $\lambda_3(\hat{\mathbf{p}}_B)$ and $\lambda_4(\mathbf{x})$ enforce fixed dipole magnitude ($\dot{\hat{\mathbf{p}}}_B\cdot\hat{\mathbf{p}}_B=0$) and continuum level incompressibility ($\nabla\cdot\mathbf{v}=0$), respectively.

The flow-induced rate of the dipole orientation of a Brownian particle is $\dot{\hat{\mathbf{p}}}_{f,B}\!=\!\hat{\hat{\mathbf{p}}}^{\perp}_B \! \cdot \! \left( \mathbf{R} \!+\! \chi \mathbf{D} \right) \! \cdot \! \hat{\mathbf{p}}_B $, where, $\mathbf{R}$ and $\mathbf{D}$ are the antisymmetric and symmetric parts of the velocity gradient tensor $\boldsymbol{\kappa}$, respectively, and $\hat{\hat{\mathbf{p}}}^{\perp}_B \!=\!\mathbf{I}\!-\!\hat{\mathbf{p}}_B\hat{\mathbf{p}}_B$ is the transverse projection operator \citep{kim2013microhydrodynamics}. The shape factor $\chi\!\!=\!\!(\!\Theta_B^{\,2}\!-\!1\!)/(\!\Theta_B^{\,2}\!+\!1\!)$ reduces to $\chi\!\approx \! \delta_B$ in the small anisotropy limit. The dissipation function $\Phi$ ( Eq. \eqref{eq:Dissip_Phi}) accounts for molecular lattice viscous contribution ($\eta_f$) to ambient flow and hydrodynamic torques on Brownian particle arising from relative angular motion and anisotropic extensional coupling, respectively. Here, for weak anisotropy, the torque–flow coupling coefficient based on resistance tensors is $\overline{\chi} \simeq \delta_B/2$ \citep{kim2013microhydrodynamics,satoh2003introduction}.

The unified hydrodynamics across scales follow from the Rayleighian variational principle. Introducing a characteristic velocity-gradient magnitude $\dot{\gamma}$ (identified with the shear rate for general shear flows), we can write, $\left( \mathbf{R}+ \chi \mathbf{D} \right)=\dot{\gamma}(\boldsymbol{\kappa}(1+\chi)-\boldsymbol{\kappa}^T(1-\chi))$. 
Dimensionless numbers $\alpha_{2B}=p_{0B}E_0\beta$, and  $\alpha_3=\dot{\gamma}/D_{r,B}$ with the rotational diffusivity of Brownian particle model ($D_{r,B}\sim R_{B}^{-3}$) indicate magnitudes of the electric and flow field influences on the Brownian particle dynamics.
Imposing the conservation of orientation probability $\dot{f}_B =-\nabla_{\hat{\mathbf{p}}_B} \cdot (f_B \dot{\hat{\mathbf{p}}}_B)$ (with notation $\nabla_{\hat{\mathbf{p}}_B} =\partial/\partial {\hat{\mathbf{p}}_B}$) in the weak flow limit $\alpha_3\ll 1$, Onsager's principle $\delta\mathcal{R}/\delta \dot{\hat{\mathbf{p}}}_B=0$ reduces to $ f_{0B}=(\alpha_{2B} /sinh(\alpha_{2B})) e^{\alpha_{2B} \hat{\mathbf{p}}_B \cdot \tilde{\mathbf{E}}}(1+\mathcal{O}(\alpha_3))$, and hence $\rho_s \approx -\nabla \cdot (\mathbf{P}(1+\alpha_1 \alpha_2/\alpha_{2B}))$, for small $\alpha_2$ and $\alpha_{2B}$, as detailed in Appendix A.
 
Variation of the Rayleighian with respect to the velocity field, $\delta \mathcal{R}/\delta \mathbf{v}=0$, yields the Stokes equation \citep{doi2011onsager}. Upon evaluating the required orientation-weighted integrals involving $f,f_{0B}$ and following mathematical simplifications for the case of simple shear, at small $\alpha_2$ and $\alpha_{2B}$ one obtains the following with VE and ES terms as shear and isotropic stress corrections (see Appendix B).
{\setlength{\belowdisplayskip}{6pt}
\setlength{\abovedisplayskip}{6pt}
\begin{align}
\!\!\!  &  \Bigg[\!\!-\!\nabla\! \cdot\!( \eta_0\nabla\mathbf{v}\!)\!- \!\nabla\!\! \cdot \! \!\left(\! \frac{c_B \zeta_B^r \overline{\chi}\alpha_{2B}^2}{315}\tilde{E}^2  \nabla\mathbf{v} \!\right)\!\!\Bigg]\!
  + \!\Bigg[ \rho_{i}\! \nabla \psi\!
     +\! \nabla \!(\mathbf{P}\!\cdot \!\nabla \psi\!) \nonumber \\
\!\!\!   &\!  +\!\frac{\alpha_2^2}{6} \! \nabla\! \left(\!  c_s\tilde{E}^2\!\right)\!\!\Bigg]\! -\!  \nabla\!\cdot \!\Bigg[\! \lambda_4 \!+c_L  \lambda_1\!+\! \frac{4\Theta_B\alpha_{2B}^2}{15\beta}  \!\! \left(\!c_B \tilde{E}^2\!\right)\!\!\Bigg]\!\mathbf{I}\! \!= \!0 \label{eq:Stokes_yx}
\end{align}
}
Here, the three square-bracketed groups of terms correspond to shear, body force and pressure, respectively.

Molecular-scale contributions from $\delta\dot{\mathcal{F}}/\delta\mathbf{v}$ enter as a Maxwell body force, comprising Coulombic and dielectrophoretic components, together with solvent dipolar orientation variation dependent osmotic force, respectively. A pressure correction additionally emerges from the Lagrange multiplier $\lambda_1$, enforcing the lattice-level constraint on molecular compressibility, thereby renormalizing the continuum pressure $\lambda_4$. 

The HB-network contribution generates multiple stress components. The extensional–rotational coupling in the dissipation function $\Phi$ (last term of Eq. \ref{eq:Dissip_Phi}) produces a shear stress contributing along with the ambient fluid dissipation (first term of Eq. \ref{eq:Dissip_Phi}). Put together, the leading-order term yields the zero-field viscosity, $\eta_0=\eta_f+c_B\zeta_B^r \overline{\chi}/15$, while higher-order term $\mathcal{O}(\alpha_{2B})$ give rise to the viscoelectric correction. In addition, dissipation associated with relative angular motion (second term of Eq. \ref{eq:Dissip_Phi}) for the anisotropic (slender) component in a deformational flow produces an isotropic stress, recognized as a contribution to the well-known electrostrictive pressure ($\Pi_{es}$); While the spherical rotational component generates no stress.

\begin{figure}[b]
\vspace{-10pt}
\centering
\begin{tikzpicture}
\sbox0{\includegraphics[width=0.2\textwidth]{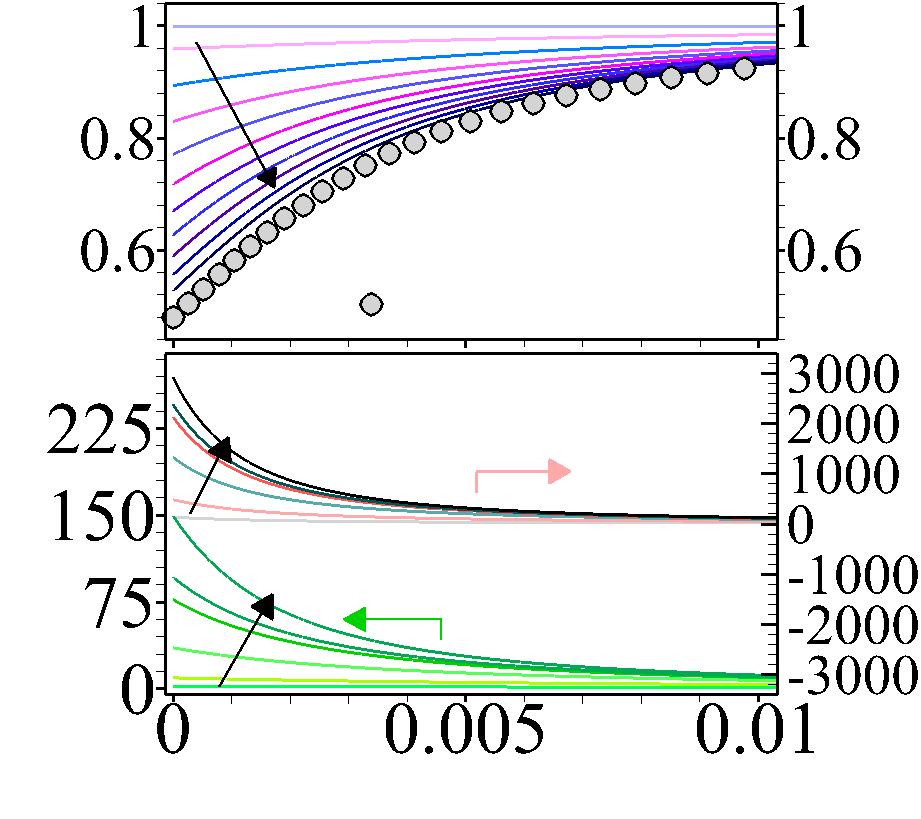} } 
  \node[above right,inner sep=0pt] at (0,0)  {\usebox{0}};
  \node[black] at (-0.05\wd0,1.0\ht0) {(a)};
    \node[black] at (0.51\wd0,0.02\ht0) {$d/L$};
  \node[black,rotate=90] at (0.01\wd0,0.75\ht0) { $\varepsilon_r$/$\varepsilon_{r0}$ };
  \node[Blue] at (0.56\wd0,0.75\ht0) { dPNP};
  \node[gray] at (0.64\wd0,0.64\ht0) { LB model};
  \node[black,rotate=90] at (-0.01\wd0,0.35\ht0) { $f_v$/$f_{v0}$ };
  \node[gray] at (0.6\wd0,0.48\ht0) { LBFT-VE};
  \node[ForestGreen] at (0.45\wd0,0.3\ht0) {dPNP-S};
\end{tikzpicture}
\begin{tikzpicture}
\sbox0{\includegraphics[width=0.2\textwidth]{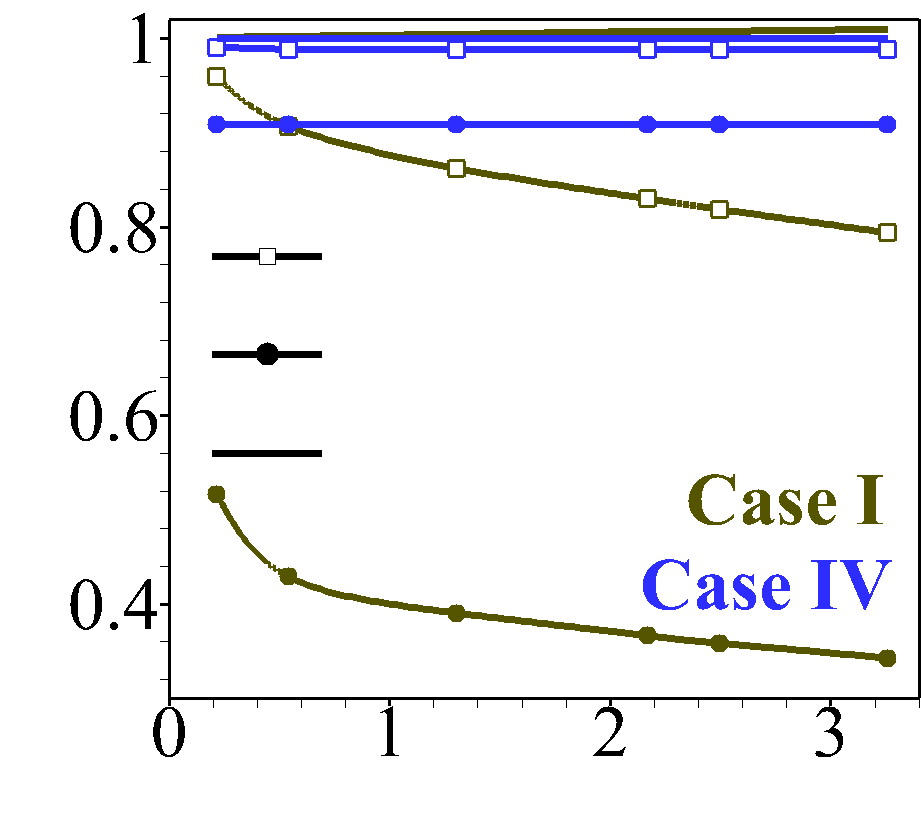} } 
  \node[above right,inner sep=0pt] at (0,0)  {\usebox{0}};
  \node[black] at (-0.05\wd0,1.0\ht0) {(b)};
  \node[black] at (0.57\wd0,0.02\ht0) {$t/\tau_D$};
  \node[black,rotate=90] at (0.01\wd0,0.57\ht0) {$\hat{\mu}_{eo}$/$\hat{\mu}_{eo,PNP}$};
    \node[black] at (0.52\wd0,0.7\ht0) { dPNP-S};
    \node[black] at (0.57\wd0,0.58\ht0) { LBFT-VE};
    \node[black] at (0.51\wd0,0.46\ht0) { mPNP};
\end{tikzpicture}
\caption{(a) Evolution of  $\varepsilon_r/\varepsilon_{r0}$ and $f_v/f_{v0}$ from dPNP-S compared to LB-fitted transient viscoelastic models (LBFT-VE). Black arrows indicate the evolution from time $0.5$ to $5$ times of Debye timescale ($\tau_D$). (b) Electroosmotic mobility correction factor over PNP for different parametric cases (refer to the \textit{joint PRF submission}), using different models. (Color online.)
} \label{fig:2_results}
\end{figure}
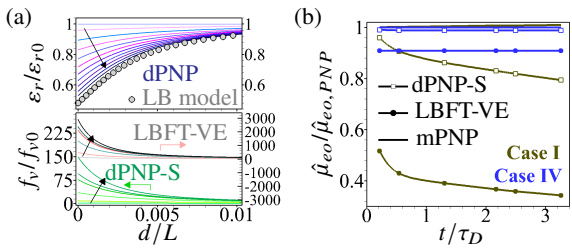

\textit{Application}-- Eq.~\eqref{eq:Poisson},~\eqref{eq:NP} and~\eqref{eq:Stokes_yx} comprise dipolar Poisson–Nernst–Planck–Stokes (dPNP–S) framework, applicable to a broad range of practical scenarios. 
The standard spatiotemporal anisotropic nonlinear relative permittivity is defined in terms of polarization density $\mathbf{P}$ as,  $\varepsilon_r\!=\!\mathbf{I}\!+\!\varepsilon_0^{-1} \nabla_\mathbf{E} \mathbf{P}$ (with $ (\nabla_\mathbf{E} \mathbf{P})_{ij} \!=\! \partial P_i/\partial E_j$ in index notation) \citep{landau2013electrodynamics}. In the limits  $\alpha_2\!\ll \!1$, $\alpha_1\ll1$, the leading order isotropic approximation $\varepsilon_r\!\approx\!\varepsilon_{r0}\!=\!1+c_s\gamma_cp_0^2 \beta \varepsilon_0^{-1}/3$ yields the field-independent estimate implying $p_0\approx |\mathbf{p}_e|$. 
Enforcing the lattice-occupation constraint on the species fluxes yields $\nabla \lambda_1\!= \!-(\sum_{i=\pm,s}\!D_ic_i\nabla M_i)/\sum_{i=\pm,s}\!D_ic_i$. For the equal size of molecules occupying the uniform lattice, the species diffusivities $D_{\pm,s}$ are comparable in magnitudes, which is a reasonable approximate for typical monovalent electrolytes. For dilute-electrolytes ($c_0/c_L\!\!\ll\!\! 1$) the weak dipolar effect ($\alpha_2\ll1$), implies $\lambda_1\!\!\rightarrow \!\!-M_s$, so that the effective solvent chemical potential in Eq. \eqref{eq:NP} vanishes, $\mu_s\!\rightarrow \!\!0$, and the molecular pressure correction in Eq: \eqref{eq:Stokes_yx} approaches: $c_L\nabla \lambda_2\!\!\rightarrow  \!c_L \alpha_2 \mathcal{L}\nabla \tilde{E}$. In appropriate asymptotic limits, the dPNP–S model recovers several established descriptions: the MLB model \citep{gongadze2012decrease} at steady state and further the LB model for $\alpha_2\ll 1, \gamma_c=1$ with appropriate $|\mathbf{p_e}|$; the modified PNP (mPNP) model \citep{kilic2007bsteric}  for $\alpha_2 \epsilon_D^2\ll1$, and further, the classical PNP model for $c_0/c_{s0}\ll 1$; and the macroscopic Stokes equation with Maxwell stress for $\alpha_1\ll 1, \alpha_2 \ll 1,\alpha_{2B} \ll 1 $. Here, $\epsilon_D\!=\!\lambda_D/L$ is the dimensionless Debye length.

The empirical viscoelectric coefficient for quadratic field dependence, estimated using the nominal electric field $E_0$ and Poisson–Boltzmann theory for the experimental conditions of \citet{jin2022direct}, is obtained from Eq.~\ref{eq:Stokes_yx} (see Appendix C) as $f_v=(0.96\pm0.4)\times10^{-15}$ (m/V)$^2$, in quantitative agreement with the reported value. Similarly, the prefactor of the electrostrictive pressure arising from HB network effects is of comparable magnitude, though slightly larger, than the statistical Kirkwood correlation factor $g_k$ (see Appendix C).

The dPNP–S equations are solved numerically for electroosmotic flow in a nanofluidic electrolytic cell with surface potential, Stern layer, and no-slip boundaries, using experimentally relevant parameters in the limit $\alpha_1 \ll 1$. 
Hydrogen-bond (HB) network dynamics introduce spatiotemporal variations in the effective permittivity and viscoelectric (VE) coefficient during electric double-layer evolution, differing from simpler transient implementations based on a constant empirical VE coefficient (denoted \textit{LBFT–VE} in the \textit{joint PRF submission}). To isolate these effects, dPNP–S predictions obtained by neglecting the electrostrictive pressure and dipolar orientational osmotic contributions in Eq.~\ref{eq:Stokes_yx} are compared with the LBFT–VE model.
Figure~\ref{fig:2_results}(a) shows that HB-network dynamics produce pronounced deviations in the evolution of the normalized permittivity $\varepsilon_r/\varepsilon_{r0}$ and VE coefficient $f_v/f_{v0}$ near the electrode surface. These variations propagate into measurable corrections in the electroosmotic mobility of the induced flow, $\mu_{eo}=\int_0^l dx\, v/(lE_0)$, relative to the classical PNP–Stokes prediction, as shown in Fig.~\ref{fig:2_results}(b) for representative parameter sets (Cases I and IV of the \textit{joint PRF submission}).For weak dipolar coupling (Case IV, $\alpha_2\epsilon_D^2\!\sim\!10^{-5}$), dPNP–S converges to mPNP, while for stronger coupling (Case I, $\alpha_2\epsilon_D^2\!=\!0.37$) it departs significantly. In both regimes, it differs from LBFT–VE, demonstrating the hydrodynamic impact of HB-network dynamics.

\textit{Insights}-- The weak anisotropy of the Brownian-particle representation ($\delta_B\ll 1 $) reflects the strong resistance of the hydrogen-bond (HB) network to shear deformation.
Unlike prior treatments assuming a constant viscoelectric coefficient $f_v$, the present theory resolves its intrinsic spatiotemporal dependence on ion concentration and electric-field strength, allowing direct implementation in continuum models.
Electrostrictive pressure macroscopically associated with dielectric variations under strain is correlated to the underlying HB network dynamics. Fundamentally, this theory provides a molecular–to–continuum multiscale connection between HB-network structure, its characteristic scales, and the resulting observable phenomena. 
Future nanofluidic measurements of EOF mobility under periodic electric fields, combined with molecular dynamics simulations, could help validate and refine the model.

\textit{Conclusion}-- To represent the collective dynamics of the hydrogen-bond network, transient clusters are modeled as orientable Brownian particles, consistent with molecular dynamics and experimental observations \citep{jansson2010hidden,zong2016viscosity,elton2017origin,gao2024structural}. Within Onsager’s variational framework, this coarse-grained description leads to a dipolar Poisson–Nernst–Planck–Stokes (dPNP–S) continuum theory for coupled ion and fluid transport. In this formulation, the hydrogen-bond network contributes as a particle-like component distinct from the background solvent response, represented by orientable dipolar entities whose rotational dynamics generate stresses reminiscent of dipolar suspensions. The resulting framework quantitatively predicts transient dielectric saturation, viscoelectric responses, and electrostrictive contributions, enabling a physically grounded solvent-enriched continuum electrohydrodynamics beyond empirical closures.

\vspace{3pt}
\textit{\textbf{Declarations}--} No data were created or used in this study. This research received no specific grant from any funding agency. The author declares no competing interests.
\bibliography{apssamp}
\vspace{6pt}
\textbf{$\quad \quad \quad \quad\quad \quad\quad \quad  $ END MATTER }
\vspace{3pt}
\appendix

\textit{Appendix A.}--  The Onsager's principle $\delta\mathcal{R}/\delta \dot{\hat{\mathbf{p}}}_B=0$ reduces to the Fokker-Planck type equation for the re-orientation dynamics of the Brownian clusters $f_B$. 
{\setlength{\belowdisplayskip}{3pt}
\setlength{\abovedisplayskip}{3pt}
\begin{align}
\!\!\dot{f}_B \!= \!\nabla_{\hat{\mathbf{p}}_B}\!\cdot\!\left( \!D_r\!\left(\! \gamma_c\beta p_{0B} f_B \nabla \psi\!  +\!\nabla_{\hat{\mathbf{p}}_B}f_B \right) \!\cdot \!\hat{\hat{\mathbf{p}}}^{\perp}_B\! -\!f _B\dot{\hat{\mathbf{p}}}_{f,B} \!\right)
\end{align}
}
For the dilute-mild aqueous electrolytes ($c_0\sim 0.01-1$ mM) the EDL evolves at micro to submicro second time scale $\tau_{EDL}=\lambda_DL/D$ depending on the translational diffusivity of ions $D_{\pm,s}\sim 1/(6\pi\eta_0\beta a)$ with lattice element size $a=c_L^{-1/3}$ and Debye length $\lambda_D=\sqrt{\varepsilon_0 \varepsilon_{r0}/(2c_0z^2e_0\beta)}\sim10-100$ nm. Hence the Brownian cluster rotation can be considered quasistatic during EDL evolution. Integrating the above equation in Einstein's indices ($i,j,k$) format, we get the following with an integration constant $c_1$.
{\setlength{\belowdisplayskip}{3pt}
\setlength{\abovedisplayskip}{3pt}
\begin{align}
   &\!\!\!\! f_B\!= c_1 e^{\frac{\alpha_3}{2}\tilde{\kappa}_{jk} \hat{p}_{Bj} \hat{p}_{Bk}-\alpha_{2B} \hat{p}_{Bi} \partial_i \psi} \nonumber\\
    &\quad \int_0^\pi \!\!\!\!d\theta sin\theta e^{(\alpha_{2B} cos\theta \,|\partial_i \psi|+\Lambda)} \!I_0\!\left(\nu_1 \right)\! I_0\!\left(\nu_2 \right)\! I_0\!\left(\!\frac{\nu_3}{2}\!\right)
\end{align}}
Here $I_0$ is the modified Bessel's function of first-kind in terms of functions of the angle $\theta$ between $\hat{\mathbf{p}}_B$ and $\mathbf{E}$,  $\nu_1=(\tilde{\kappa}_{xy}+\tilde{\kappa}_{yx})\frac{\alpha_3}{2} cos\theta\, sin\theta$, $\nu_2=(\tilde{\kappa}_{xz}+\tilde{\kappa}_{zx})\frac{\alpha_3}{2} cos\theta\, sin\theta$ and $\nu_3=(\tilde{\kappa}_{yz}+\tilde{\kappa}_{zy})\frac{\alpha_3}{2} sin^2\theta$. $\Lambda=\tilde{\kappa}_{xx} cos^2(\theta) +
\tilde{\kappa}_{yy} sin^2(\theta)cos^2(\phi) +
\tilde{\kappa}_{zz} sin^2(\theta)sin^2(\phi)$ consists of only the extensional terms of the velocity gradient.
For a general shear flow (i.e., $\Lambda=0$) with small advection $\alpha_3 \ll 1$, this approximates using the condition $\left<f_B\right>=1$ to: 
{\setlength{\belowdisplayskip}{3pt}
\setlength{\abovedisplayskip}{3pt}
\begin{align}
    f_B=& f_{0B}  \left(\! 1\!+\! \alpha_3 (2\overline{\chi}\hat{\mathbf{p}}_B \! \cdot \! \mathbf{D} \cdot \hat{\mathbf{p}}_B) \! +\! \mathcal{O}(\alpha^2) \right),\\
    f_{0B}=&(\alpha_{2B} e^{\alpha_{2B} \hat{\mathbf{p}}_B \cdot \tilde{\mathbf{E}}})/sinh(\alpha_{2B}).\nonumber
\end{align}}

\textit{Appendix B.}-- Useful Integrals: 
since $f_0$ is axisymmetric about the perturbing electric field $\mathbf{E}$, any tensorial moment of $f_0$ shall be constructed from the available invariant tensors under rotation of $\mathbf{E}$, such as $\delta_{ij},\hat{E}_i\hat{E}_j$,$\delta_{ij}\hat{E}_k\hat{E}_l$, $\hat{E}_i\hat{E}_j\hat{E}_k\hat{E}_l$. Integrals below are evaluated using such an ansatz, when required.
{\setlength{\belowdisplayskip}{3pt}
\setlength{\abovedisplayskip}{3pt}
\begin{align}
  &\!\!\!\!\!\frac{\mathbf{I}_1}{4\pi}\!\!=\!\!\frac{\int \!\!\!d\hat{\mathbf{p}}_B f_0 \hat{\mathbf{p}}_B}{4\pi}\! \!= \! \!\mathcal{L}_B \hat{\mathbf{E}},\, \mathcal{L}_B\!=\! coth(\alpha_{2B}\!\tilde{E})\!-\!(\!\alpha_{2B} \tilde{E})\!^{-\!1} \!
  \end{align}}
{\setlength{\belowdisplayskip}{3pt}
\setlength{\abovedisplayskip}{3pt}
  \begin{align}
  &\!\!\!\!\!\!\int\!\! \!\!d\hat{\mathbf{p}}_B f_0\! \ln(\!f_0)\! =  \!4 \pi\! \left( \!\ln\! \left(\!\alpha_{2B}\tilde{E}/sinh(\!\alpha_{2B}\tilde{E} )\right)\!\! +  \!\alpha_{2B}\tilde{E} \! \mathcal{L}_B\! \right) \!\!
  \end{align}
} {\setlength{\belowdisplayskip}{3pt} \setlength{\abovedisplayskip}{3pt}
  \begin{align}
&\!\!\!\!\int\!\! \!\!d\hat{\mathbf{p}}_B  f_0 (\!- \hat{\mathbf{p}}_B\! \cdot \!\nabla \psi\!)^k \! =\! \frac{4\pi}{\alpha_{2B}^k Z}\!\dv[k]{Z}{\tilde{E}}, \, Z\! =\!\frac{2\! sinh(\!\alpha_{2B} \tilde{E} )}{\alpha_{2B} \tilde{E}}
\end{align}
}{\setlength{\belowdisplayskip}{3pt} \setlength{\abovedisplayskip}{3pt}
  \begin{align}
&\!\!\!\mathbf{I}_2\!=\!\!\!\int\!\!\!\! d\hat{\mathbf{p}}_B f_0\! \left(\!\hat{\mathbf{p}}_B \hat{\mathbf{p}}_B\! - \! \frac{\mathbf{I}}{3} \!\!\right)\!\! = \!4\pi \!\!\left( \!\left(  \frac{\mathcal{L}_B}{\!\alpha_{2B} \tilde{E}}\! -\!\frac{1}{3} \!\right) \!\mathbf{I} \!\!+\!\! \left(\!1\!\!-\! \frac{3\mathcal{L}_B}{\alpha_{2B} \tilde{E}}\!\right) \!\hat{\mathbf{E}}\hat{\mathbf{E}} \!\right),\nonumber
\end{align}
} {\setlength{\belowdisplayskip}{3pt} \setlength{\abovedisplayskip}{3pt}
  \begin{align}
&\!\!\! \mathbf{I}_3\!=\! \!\int \!\!\!d\hat{\mathbf{p}}_B\,f_0 \hat{\mathbf{p}}_B\hat{\mathbf{p}}_B(\hat{\mathbf{p}}_B \cdot \nabla \psi) \nonumber\\
&= \! -4 \pi|\nabla \psi| \left[\!\left(\!\left(\!1\!+\! 15/(\alpha_{2B}\tilde{E})^2\!\right)\!\mathcal{L}_B\!-\!5/(\alpha_{2B}\tilde{E}) \!\right) \hat{\mathbf{E}}\hat{\mathbf{E}}\right.\! \nonumber\\
&\quad  \!\left.\! + \!\left(\!1/(\alpha_{2B}\tilde{E})\!-\!3\mathcal{L}_B/(\alpha_{2B}\tilde{E})^2 \!\right)\! \!\left(\!\hat{\mathbf{E}} \hat{\mathbf{E}}+(\hat{\mathbf{E}} \hat{\mathbf{E}})^T\!\! -\!\mathbf{I}\!\right)\! \right]
\end{align}}  {\setlength{\belowdisplayskip}{3pt} \setlength{\abovedisplayskip}{3pt}
  \begin{align}
&\!\!\!\!\int \!\! \!\!d\hat{\mathbf{p}}_B\,f_0 \hat{\mathbf{p}}_B(\hat{\mathbf{p}}_B \cdot \nabla \psi)= 4\pi \nabla \psi \left(1-2\mathcal{L}_B/(\alpha_{2B}\tilde{E}) \right)
\end{align}
}  {\setlength{\belowdisplayskip}{3pt} \setlength{\abovedisplayskip}{3pt}
  \begin{align}
&\!\!\!\!\!\int \!\! \!\!d\hat{\mathbf{p}}_B f_0 \hat{\mathbf{p}}_B\!(\!\hat{\mathbf{p}}_B \cdot \nabla \psi\!)\!=\! 4\pi \nabla \psi \!\left(\!1\!-\!2\mathcal{L}_B/(\alpha_{2B}\tilde{E})\! \right)
\end{align}
} Also, define a symmetric tensor $\mathbf{I}_4\!=\!p_{0B}\beta (\mathbf{I}_1\! \nabla   \psi\!-\! \mathbf{I}_3 )\!+\!3\mathbf{I}_2=-8 \pi\mathbf{I}(1-3\mathcal{L}_B/\alpha_{2B}\tilde{E})$. 

Using these integrals, we get $\delta \Phi/\delta \mathbf{v}= - \nabla \cdot \left(  \eta_f \bm{\kappa}+\bm{\sigma_1}+\bm{\sigma_2}+\bm{\sigma_3} \right)$ with $4\pi\bm{\sigma_1}=-c_B (\mathbf{I}_{4}-\mathbf{I}_{4}^T)=0$, $4\pi \bm{\sigma_2}=\chi c_B(\mathbf{I}_{4}+\mathbf{I}_{4}^T)$. For a small HB energetic effect, $\alpha_{2B}\ll 1$, approximating $3\mathcal{L}_B/\alpha_{2B}\tilde{E}=1-\alpha_{2B}^2/15+ \mathcal{O}(\alpha_{2B}^4)$, the isotropic stress $\bm{\sigma}_2=4\chi c_B\alpha_{2B}^2\tilde{E}^2\mathbf{I}/15\beta$, recovers a local field-dependent pressure correction, which is recognized with the classical electrostrictive pressure. Similarly, the stress component, 
 {\setlength{\belowdisplayskip}{3pt} \setlength{\abovedisplayskip}{3pt}
\begin{align}
&\!\!\bm{\sigma}_3\!= c_B \zeta_B^r \frac{\overline{\chi}}{4\pi} \left[ C_1(\hat{\mathbf{E}} \cdot \mathbf{\kappa} \cdot \hat{\mathbf{E}})\hat{\mathbf{E}}\hat{\mathbf{E}}\right.\nonumber\\
    &\quad+ C_2( Tr(\mathbf{\kappa})\,\hat{\mathbf{E}}\hat{\mathbf{E}} 
    + \hat{\mathbf{E}} \hat{\mathbf{E}} \cdot \mathbf{\kappa} 
    +( \mathbf{\kappa} \cdot \hat{\mathbf{E}}  \hat{\mathbf{E}})^T 
    + \hat{\mathbf{E}} \cdot \mathbf{\kappa} \hat{\mathbf{E}}  
    + \mathbf{\kappa} \cdot \hat{\mathbf{E}} \hat{\mathbf{E}}\nonumber\\
    &\quad + \hat{\mathbf{E}} \cdot \mathbf{\kappa} \cdot \hat{\mathbf{E}} \mathbf{I}) \left. + C_3 (Tr(\mathbf{\kappa})\mathbf{I} +\mathbf{\kappa} +(\mathbf{\kappa})^T) \right], \nonumber 
\end{align}}
 
Also, another integral $\mathbf{I}_5\!=\!\int \!\!d\hat{\mathbf{p}}_B\,f_0  \hat{p}_i  \hat{p}_j  \hat{p}_\mu  \hat{p}_\nu $ is useful for evaluating $8\pi \bm{\sigma_3}= c_B \zeta_B^r\overline{\chi}(\bm{\kappa}+\bm{\kappa}^T) \mathbf{:} \mathbf{I}_5$. Owing to the orientation probability about $\hat{\mathbf{E}}$, this integral can be expanded in fourth order moments $\boldsymbol{\Gamma}_1 = 
\hat{\mathbf E}^{\otimes 4},
\boldsymbol{\Gamma}_2= 
\hat{\mathbf E}\hat{\mathbf E}\otimes\mathbf I
+\mathbf I\otimes\hat{\mathbf E}\hat{\mathbf E}
+\mathcal{P}(\hat{\mathbf E}\otimes\mathbf I\otimes\hat{\mathbf E})$ and $ \boldsymbol{\Gamma}_3 = 
\mathbf I\otimes\mathbf I
+\mathcal{P}(\mathbf I\otimes\mathbf I)$ (where $\otimes$ and  $\mathcal{P}$ represents tensor products, and indices permutations, respectively) as, $\mathbf{I}_5\!=\!\sum_i C_i\mathbf{\Gamma}_i$. The coefficients are identified using contraction of $\mathbf{I}_5$ with $\mathbf{\Gamma}_i$ as, $C_i\!= \frac{\pi}{2}\left(
Q_i + \frac{U_i}{(\alpha_{2B}\tilde{E})^2}
-\left(
\frac{V_i}{\alpha_{2B}\tilde{E}}
+\frac{W_i}{(\alpha_{2B}\tilde{E})^3}
\right)\mathcal{L}_B
\right)$, with $(Q_i,U_i,V_i,W_i)= (8,280,80,840), (0,-40,-8,-120)$, $(0,8,0,24)$ for $i=1,2,3$ respectively.
The case of simple shear flow (say, having $\kappa_{yx}=\dot{\gamma}$, while other components are zero) perpendicular to the field ($\mathbf{E}=E\hat{\mathbf{x}}$) can be used to simplify the stress to $\bm{\sigma}_3=\dot{\gamma}(C_2+C_3)$, which is $\sigma_{yx}=c_B\zeta^r_B\overline{\chi}\dot{\gamma}(1+\alpha_{2B}^2\tilde{E}^2/21)/15$ for $\alpha_{2B}\ll1$. Hence, this contributes to the field-dependent strain rate-based viscous shear stress.

\vspace{5pt}
\textit{Appendix C.}--\citet{jin2022direct}'s experimental conditions include $T=298\text{ K}, c_0\!=\!0.08 \text{ mM}, E_0\!=\!(\!V_1\!-\!V_2\!)/L, V_1\!=\!-150 \text{ mV},V_2\!=\!100 \text{ mV}, L\!=\!57.5\pm17.5 \text{ nm}$, while using PB theory. Hence, $f_{v0}=L_E L_m f_{vB}$, $f_{vB}={c_B\zeta_r \overline{\chi} \beta^2 p_{0B}^2}/{315 \eta_0}$ comes from the HB network segment, while $L_E=(E_s/E_0)^2$ and $L_m= \left( \frac{(1-A_1E_s^2-A_2 \psi_s^2)\epsilon_D(1-e^{-1/\epsilon_D})}{1-(A_1E_s^2-A_2 \psi_s^2)\epsilon_D(1-e^{-1/\epsilon_D})} \right)$ are prefactors for the nominal field to surface field and surface value of  to mean value conversions. 
Here $E_s(c_0,V_0,L)$ is the electric field at the electrode surface of constant potential, estimated iteratively, and $A_i$ are the coefficients from MLB model \citep{gongadze2012decrease}, $A_1A_3= 2c_{s0}\alpha_2^4/(135\varepsilon_0E^2\beta)$ and $A_2A_3= 2c_0\alpha_2^2(e_0\beta\psi)^2/(9 \varepsilon_0 E^2\beta)$, with $A_1 = n^2+ 2\alpha_2^2 c_{s0}/(9\varepsilon_0\beta E^2)$.
Similarly, electrostrictive pressure $\Pi_{es}= L_p \,\Pi_{0,es}$, with the traditional expression $\Pi_{0,es}\! = \!c_{s0}|\mathbf{p_e}|^2\beta E^2/6$ and the HB network prefactor $L_p\!= \!\left(\frac{8}{5} \Theta_B \alpha_1^2 n_B \!\right)\! \left(\!\frac{c_s \overline{p}_0^2}{c_{s0} |\mathbf{p_e}|^2}\! \right)$.

\end{document}